\documentclass[journal, letterpaper, 10pt]{IEEEtran}
\usepackage{amssymb}
\usepackage{amsthm}
\usepackage{latexsym}
\usepackage[noadjust]{cite}

\usepackage{color}

\usepackage[font={footnotesize}]{caption}
\usepackage[font={footnotesize}]{subcaption}
\usepackage{csquotes}
\usepackage{hyperref}
\usepackage{url}
\usepackage{epsf}
\usepackage{dcolumn}
\usepackage{float}
\usepackage{lipsum}
\usepackage{widetext}
\usepackage{cuted} 
\usepackage{textcomp}
\usepackage{algorithm}
\usepackage{algorithmicx}
\usepackage{algpseudocode}

\usepackage{wrapfig}
\usepackage{tikz}
\usetikzlibrary {arrows.meta}
\usepackage{graphicx}
\usepackage{amsmath}
\usepackage[cmintegrals]{newtxmath}
\theoremstyle{definition}

\theoremstyle{definition}

\theoremstyle{remark}

\newcommand{\argmax}{arg~max_}

\begin{document}
%
\title{Implementing QZMAC (a Decentralized Delay Optimal MAC) over 6TiSCH under the Contiki OS in an IEEE 802.15.4 Network}

%
%
%

\author{ Shivam~Vinayak~Vatsa\textsuperscript{1},
         Avinash~Mohan\textsuperscript{2},
        and~Anurag~Kumar\textsuperscript{1},\\
        \textsuperscript{1}Indian Institute of Science, Bangalore, India.\\
        \textsuperscript{2}Technion, Israel Institute of Technology, Haifa, Israel.

}
\maketitle

\begin{abstract}
Motivated by the emerging delay-sensitive applications of the Internet of Things (IoT), there has been a resurgence of interest in developing medium access control (MAC) protocols in a time-slotted framework. The resource-constrained, ad-hoc nature of wireless networks typical of the IoT also forces the amount of control information exchanged across the network –- required to make scheduling decisions -– to a minimum. In a previous article we proposed a protocol called QZMAC that (i) provides provably low mean delay, (ii) has distributed control (i.e., there is no central scheduler), and (iii) does not require explicit exchange of state information or control signals. 

In the present article, we implement and demonstrate the performance of QZMAC on a test bed consisting of CC2420 based Crossbow telosB motes, running the 6TiSCH communication stack on the Contiki operating system over the 2.4GHz ISM band. QZMAC achieves its near-optimal delay performance using a clever combination of \emph{polling} and \emph{contention} modes. We demonstrate the polling and the contention modes of QZMAC separately. We use an Adaptive Synchronization Technique in our implementation which we also demonstrate. Our network shows good delay performance even in the presence of heavy interference from ambient WiFi networks.
\end{abstract}

\begin{IEEEkeywords}
Internet of Things (IoT), Medium Access Control (MAC) protocols, IEEE 802.15.4, 6TiSCH.
\end{IEEEkeywords}

%
\IEEEpeerreviewmaketitle

\section{Introduction}
\begin{figure}
\centering
\input{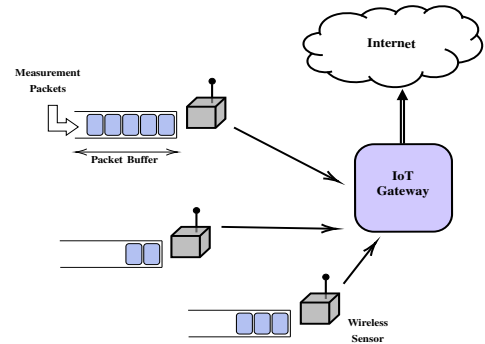}
\caption{A typical application scenario. Each of the $N$ sensors generates measurement packets that get queued in its packet buffer. These are then transmitted over a \emph{wireless} link to a gateway that forwards the packet to the broader internet, i.e., the \emph{infrastructure network.}}
\label{figWsnGatewayInternet}
\vspace{-0.60cm}
\end{figure}

In the Internet of Things (IoT), wireless access networks will connect embedded sensors to the infrastructure network (see Fig.~\ref{figWsnGatewayInternet}). Since these embedded devices will be resource challenged, the wireless medium access control (MAC) protocols will need to be simple, and decentralized, and should not require explicit exchange of state information and control signals. However, some of the emerging applications might expect low packet delivery delays as well \cite{tight-delay-ref}. 
Moreover, emerging standards for IoT applications, such as the DetNet 
and 6TiSCH \cite{thubert-etal15sdn-meets-iot}, have shown considerable interest in systems with a 
synchronous time-slotted framework. 

In an earlier article \cite{mohan-etal16hybrid-macsMASSversion}, we proposed and analyzed QZMAC, a low mean delay MAC protocol for $N$ collocated nodes sharing a time-slotted wireless channel, which requires no centralized control and no explicit exchange of state information. 
It may be noted that, for this setting, a centralised scheduler with full queue length information can just schedule any non-empty queue in each slot. The challenge, therefore, is to develop a distributed mechanism, without explicit exchange of queue length information, that achieves mean delay very close to that of the centralised scheduler. QZMAC accomplishes this through a clever combination of \emph{contention} and \emph{polling.}

It is well known that, while contention access (ALOHA, CSMA etc.) performs well at low contention, it can result in very large delays
and possibly instability under high contention \hspace{-0.05cm}\cite{lam84principles-comm-networking-prtcls}. Despite attempts having been made to stabilize CSMA (all relevant theoretical references are provided in our earlier article \cite{mohan-etal16hybrid-macsMASSversion}), the delay of these algorithms still remains prohibitively high. Polled access (e.g., 1-limited cyclic service, which we will call TDMA in this paper)
 on the other hand, shows the opposite behavior. 
It is, hence, desirable to have protocols that can behave like TDMA
under high contention and CSMA under low contention. 
There have been many attempts at proposing protocols that achieve this, 
especially in the context of wireless sensor networks. 

The design of these protocols, however, has been heuristic and without recourse to known results in optimal scheduling of queues. Further, little attention appears to have been paid to distributed scheduling in the TDMA context, such that each node uses only locally available information.  If the empty-nonempty statuses of queues are known at a central scheduler, then ideal performance can be achieved. In the context of partial information about queue occupancies, the aim, at each scheduling instant, is to minimize the time it takes for the system to find a nonempty queue and allow it to transmit and, as Sec.~\ref{sec:systemDescriptionAndQZMAC} shows, QZMAC is designed to do precisely this.

We implemented QZMAC as a new module in 6TiSCH communication stack under Contiki OS \cite{contiki-ref}. We exploit the slots architecture defined in IEEE 802.15.4-2015e TSCH and the API provided by Contiki OS. We test our implementation under COOJA simulation environment. Then we port our program to a testbed containing $5$ Crossbow telosB motes including a Border Router. Through the test bed experiment, we verified the working of polling mode, contention mode, detection of channel state using CCA and synchronization within the network.     

\section{System Description and The QZMAC Protocol}\label{sec:systemDescriptionAndQZMAC}
\begin{figure}[tb]
\centering
\includegraphics[height=3.25cm, width=7.25cm]{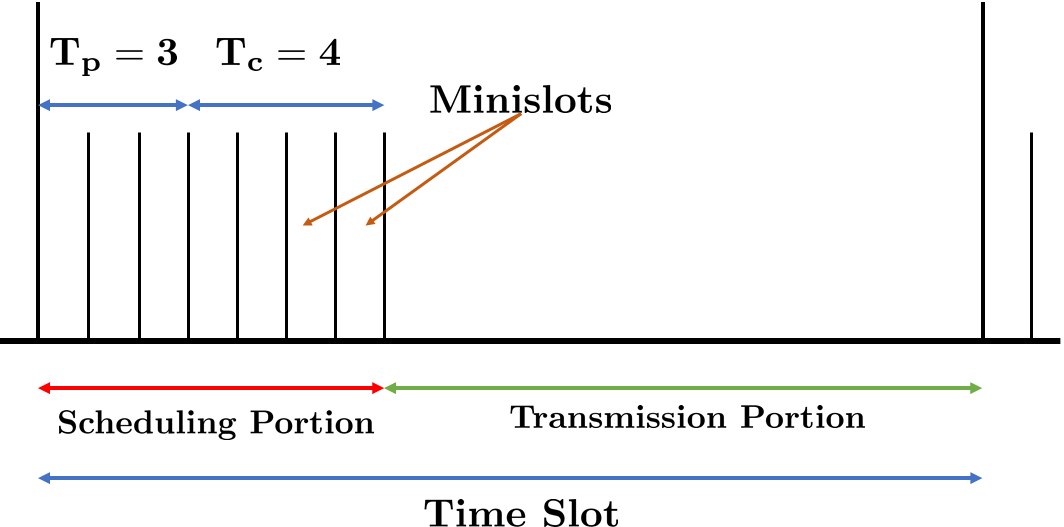}
\caption{Illustrating the slot and minislot structures. Since $T_p=3$, \emph{three} poll-and-test procedures can be performed by the system (incumbent plus two other queues). If none of these results in identifying a nonempty queue, the system goes into contention over the next $T_c=4$ minislots.}
\label{figMinislotStructure}
\vspace{-0.5cm}
\end{figure}

 We consider $N$ sensor nodes sharing a time-slotted wireless channel to transmit packets (each of which fits into one slot) to a common receiver that we call the \emph{IoT Gateway} (see Fig.~\ref{figWsnGatewayInternet}). 
 Our objective, as stated before, is to construct a protocol that (a) shows good delay performance, (b) smoothly transitions from contention access to polled access depending upon system load\footnote{By \enquote{load} we mean the total rate of packet arrival to the network.}, and (c) can be implemented in a decentralized manner. We aim to accomplish this by a combination of insights from theory and practice. We first limit ourselves to scheduling decisions based solely on buffer occupancy information gleaned from a node when it is transmitting. This obviates the need for the nodes to, for example, periodically exchange buffer information. Our optimal scheduling analysis in this setting \cite[Thm.~III.2]{mohan-etal16hybrid-macsMASSversion} shows that the \emph{polling} policy with lowest mean delay allows the currently transmitting node to continue until it is empty and then switches to the node that hasn't transmitted the longest. We therefore need only to know if nodes are empty or nonempty for scheduling. The first two of the $T_p$ minislots in Fig.~\ref{figMinislotStructure} are reserved for this.
 
 QZMAC achieves this in a decentralized manner by employing a special time slot structure. Each time slot comprises a \emph{Scheduling} and a \emph{Transmission} and the former is further subdivided into multiple minislots (Fig.~\ref{figMinislotStructure}). Each minislot is long enough for the nodes to perform a Clear Channel Assessment (CCA) to determine if there some node is transmitting over the channel. The nodes all maintain a vector $\mathbf{V}_{N\times1}$ containing the number of slots since each of the $N$ nodes transmitted. The entry corresponding to the incumbent, called the \emph{Primary User} (PU), is obviously $0.$  Furthermore, to accommodate the fact that polling with partial information might schedule an empty queue, QZMAC also possesses a \emph{contention} phase, that takes place over the $T_c$ minislots as described below. 
 At time slot $t\geq0$:
 \begin{enumerate}
     \item\label{qzmacExhaustiveServiceStep} If the PU (say $j$) is nonempty, it begins transmission in the first minislot. All queues sense this and update V(t)
as follows:
\begin{equation}
V_i(t+1) := 
\begin{cases}
0,\text{ if }i = j\\
V_i(t)+1,\text{ otherwise}
\end{cases}
\end{equation}

\item\label{qzmacArgMaxVIStep} If it is empty, the node $i^* := \argmax{1\leq i\leq N}V_i(t)$ is allowed to transmit (in the next minislot). The queues update $\mathbf{V}(t)$ as in the previous step, replacing $j$ with $i^*.$

\item\label{qzmacSUTransmssionStep} If $i^*$ is empty, the node that won the previous contention (whenever it occurred), henceforth called the \emph{Secondary User} (SU), is allowed to transmit.

\item\label{qzmacContentionStep} Finally, if this is also empty, the queues contend over the $T_c$ minislots (using a random backoff mechanism). Suppose there is no collision and an SU, say $k$, wins, then Node~$k$ will transmit one packet during this slot.
 \end{enumerate}
 \subsection{Discussion}
 To begin with, the description of QZMAC clearly shows that \emph{no explicit} information exchange is required to implement the protocol. The empty-nonempty statuses of relevant nodes are inferred from channel activity. While this certainly limits the geographical spread of a network over which QZMAC can be run (to prevent CCA errors), the decentralized nature, ease of implementation and excellent delay performance make it a very good candidate for single hop IoT applications envisioned in many upcoming 5G networks.
 
Secondly, when the system is lightly loaded, most of the packet buffers are empty and the $T_p$ polling minislots will mostly fail to identify a nonempty node. This means that QZMAC enters its contention phase (Step~\ref{qzmacContentionStep}) and behaves mostly like a contention protocol. But when the system is heavily loaded, polling mostly succeeds and QZMAC behaves like a polling protocol thus achieving the smooth, load-dependent transition from contention to polled access we required. QZMAC thus, is a legitimate hybrid MAC protocol. 

\section{Implementation}
We implemented the QZMAC algorithm as an additional module in the MAC layer of the 6TiSCH communication stack under Contiki \cite{contiki-ref}. A configurable interface is available within Contiki for interaction between MAC layer and upper layers. This interface is basically set of callback functions. The transmitter and receiver modules are implemented as separate \enquote{protothreads} \cite{prothread-ref}. The radio always remains in polling mode to receive a packet and an interrupt is generated whenever there is a transmission to take place. We used Contiki driver API to manage other miscellaneous wireless transceiver functionalities. 

\begin{figure}[tb]
    \centering
    \includegraphics[height = 4.75cm, width = 4.755cm]{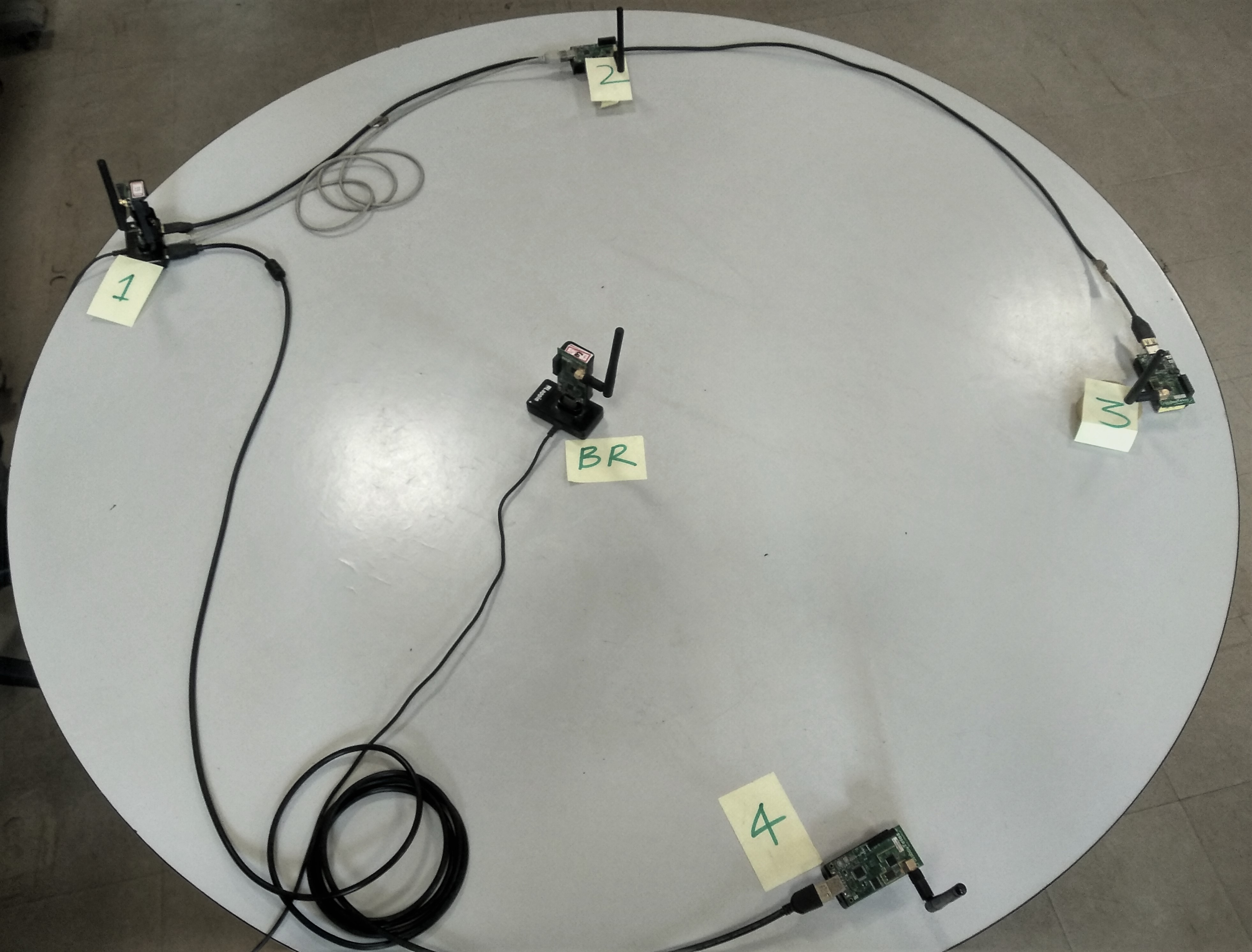}
    \caption{The QZMAC testbed setup. Here, the center node is the Border Router for the network. The other nodes, numbered 1 to 4, are the sensor nodes. These nodes are programmed to run the QZMAC protocol and transmit packets to the Border Router.}
    \label{fig:qzmac-testbed}
    \vspace{-0.50cm}
\end{figure}

We carry out our experiment over Channel~15 of the 2.4 GHz ISM band. We deployed CC2420 based telosb (sky) motes placed equidistant from the receiver node on a circular table Fig.~\ref{fig:qzmac-testbed}. Here, the node placed in the center acts as a \enquote{Border Router} (BR). The BR is always connected to a PC (host) through a USB cable, collects the data from the sensor nodes and sends it to the host (corresponding to the \emph{IoT Gateway} in Fig.~\ref{figWsnGatewayInternet}) which can be further routed to the internet 
Nodes $1$ to $4$ are sensor nodes. 

The nodes form a $1-$hop fully connected network. We used the \enquote{Routing Protocol for Low-Power and Lossy Links} (RPL) to form the route within the network \cite{winter-etal12RPL} and verified the working of our firmware on the COOJA simulator \cite{cooja-ref}, before compiling it on to real target motes. We now describe our experimentation methodology in detail
\vspace{-0.75em}

\subsection{Frame Structure: Time Slots and Mini-Slots}
Time Slotted Channel Hopping (TSCH) is a MAC layer specified in IEEE 802.15.4-2015 \cite{ieee-tsch-std}, with a design inherited from WirelessHART and ISA 100.11a.In our implementation, we used the slots implemented as part of TSCH within Contiki (sans the channel hopping utility), where each slot is $10ms$ long. A time slot is sufficiently long for the transmitter to transmit the longest possible packet and for the receiver to send back an acknowledgment.
The slots are further divided into polling mini slots and contention mini slots as shown in Fig.~\ref{figMinislotStructure}.
It is not easy to keep synchronization maintained at the level of the mini slots, so we designed the mini slots using two standard Clear Channel Assessment (CCA) duration, where $1$ CCA duration is $128$ microseconds . Through our experiments, we verified that clock drifts were not affecting the protocol's working across the mini slots.
In our implementation, We used $9$ contention and $3$ polling mini slots i.e. $T_c=9$, and $T_p=3$.
\vspace{-0.75em}
\subsection{Time Synchronization}
For our experiment, we used the Adaptive Synchronisation Technique. The Border Router broadcasts Enhanced Beacons (EBs) periodically containing a field indicating the current slot number also known as Absolute Slot Number (ASN). The other nodes store the ASN value and increment it every slot to keep the time slot number aligned. To align the time slot boundaries, the first mini-slot begins after a guard time offset of {\tt TsTxOffset} from the leading edge of every slot. Every node timestamps the instant it starts receiving the EB, and then aligns its internal timers so that its slot starts exactly {\tt TsTxOffset} before the reception of EB. In our implementation we used {\tt TsTxOffset}=$1.8ms$ 
\vspace{-0.75em}
\subsection{PU, SU, and V-vector implementation}
We preprogrammed the nodes with the variables $PU$ for primary user, $SU$ for secondary user and $\mathbf{V},$ the vector indicating the number of slots that have elapsed since each node was served last. During implementation, we randomly assigned values to $PU$ and $SU$ with the id of a node in network and we initialized $\mathbf{V}$ as $V_i = i$, with $i$ being the node id of the $i$\textsuperscript{th} deployed mote. The preprograming ensures that the value of these variables are same across all the nodes and  prevents the conflicts in computing $i^\ast$ in Step~\ref{qzmacArgMaxVIStep} of QZMAC.    
\vspace{-0.75em}
\subsection{Detection of Contention Winner}
In $T_p =2$, $i^* := \argmax{1\leq i\leq N}V_i(t)$ is found, $PU$ variable is updated with $i^*$ \& node $i^*$ transmits as explained in step 2 of QZMAC and other nodes perform CCA. If the CCA is successful during $T_p =2$ minislots, the protocol enters contention mode (Step~$3$ ,$4$ in QZMAC). In this mode all non empty nodes draw a number '{\tt r}' at random  between $1$ to $9$. The nodes goes into back-off for ${\tt r}-2$ duration, then performs CCA. If CCA is successful, they transmit during the remainder of that slot. This step ensures that the node drawing minimum '{\tt r}', wins the contention and will transmit.  
The receive procedure, which is implemented as as separate protothread, intercepts the ongoing transmission and processes the packet partially to extract the id of the transmitter before discarding the packet and update the $SU$ variable as in step $4$.
If two nodes draw same '{\tt r}', then collision occurs.
\vspace{-0.75em}
\section{Conclusion}
In conclusion, our demo will feature the following items:
\begin{itemize}
  \item Working of the Contention and Polling modes
  \item CCA status inference across the slots, and
  \item Synchronization within the network 
\end{itemize}
  In the future, we plan to test the robustness of QZMAC to CCA errors, variable rate arrival processes, radio duty cycling and nodes entering and leaving the network. Another crucial question this test bed will help answer is the energy consumption of networks running QZMAC versus other hybrid MAC protocols.
\vspace{-1em}
%

\bibliographystyle{IEEEtran}
\bibliography{IEEEabrv,techreport}

\begin{thebibliography}{1}
\providecommand{\url}[1]{#1}
\csname url@rmstyle\endcsname
\providecommand{\newblock}{\relax}
\providecommand{\bibinfo}[2]{#2}
\providecommand\BIBentrySTDinterwordspacing{\spaceskip=0pt\relax}
\providecommand\BIBentryALTinterwordstretchfactor{4}
\providecommand\BIBentryALTinterwordspacing{\spaceskip=\fontdimen2\font plus
\BIBentryALTinterwordstretchfactor\fontdimen3\font minus
  \fontdimen4\font\relax}
\providecommand\BIBforeignlanguage[2]{{%
\expandafter\ifx\csname l@#1\endcsname\relax
\typeout{** WARNING: IEEEtran.bst: No hyphenation pattern has been}%
\typeout{** loaded for the language `#1'. Using the pattern for}%
\typeout{** the default language instead.}%
\else
\language=\csname l@#1\endcsname
\fi
#2}}

\bibitem{tight-delay-ref}
\BIBentryALTinterwordspacing
F.~Mager, D.~Baumann, R.~Jacob, L.~Thiele, S.~Trimpe, and M.~Zimmerling,
  ``Feedback control goes wireless: Guaranteed stability over low-power
  multi-hop networks,'' in \emph{Proceedings of the 10th ACM/IEEE International
  Conference on Cyber-Physical Systems}, ser. ICCPS '19.\hskip 1em plus 0.5em
  minus 0.4em\relax New York, NY, USA: Association for Computing Machinery,
  2019, p. 97–108. [Online]. Available:
  \url{https://doi.org/10.1145/3302509.3311046}
\BIBentrySTDinterwordspacing

\bibitem{thubert-etal15sdn-meets-iot}
P.~Thubert, M.~R. Palattella, and T.~Engel, ``6{T}i{SCH} centralized
  scheduling: when {SDN} meets {I}o{T},'' in \emph{Proc. of IEEE Conf. on
  Standards for Communications and Networking (CSCN, 2015)}, 2015.

\bibitem{mohan-etal16hybrid-macsMASSversion}
A.~Mohan, A.~Chattopadhyay, and A.~Kumar, ``Hybrid {M}{A}{C} protocols for
  low-delay scheduling,'' in \emph{Mobile Ad Hoc and Sensor Systems (MASS),
  2016 IEEE 13th International Conference on}.\hskip 1em plus 0.5em minus
  0.4em\relax IEEE, 2016, pp. 47--55.

\bibitem{lam84principles-comm-networking-prtcls}
S.~S. Lam~(editor), \emph{Principles of Communication and Networking
  Protocols}.\hskip 1em plus 0.5em minus 0.4em\relax IEEE Computer Society
  Press, 1984.

\bibitem{contiki-ref}
A.~{Dunkels}, B.~{Gronvall}, and T.~{Voigt}, ``Contiki - a lightweight and
  flexible operating system for tiny networked sensors,'' in \emph{29th Annual
  IEEE International Conference on Local Computer Networks}, Nov 2004, pp.
  455--462.

\bibitem{prothread-ref}
\BIBentryALTinterwordspacing
A.~Dunkels, O.~Schmidt, T.~Voigt, and M.~Ali, ``Protothreads: Simplifying
  event-driven programming of memory-constrained embedded systems,'' in
  \emph{Proceedings of the 4th International Conference on Embedded Networked
  Sensor Systems}, ser. SenSys '06.\hskip 1em plus 0.5em minus 0.4em\relax New
  York, NY, USA: ACM, 2006, p. 29–42. [Online]. Available:
  \url{https://doi.org/10.1145/1182807.1182811}
\BIBentrySTDinterwordspacing

\bibitem{winter-etal12RPL}
T.~Winter \emph{et~al.}, ``Rpl: Ipv6 routing protocol for low-power and lossy
  networks.'' \emph{RFC}, vol. 6550, pp. 1--157, 2012.

\bibitem{cooja-ref}
F.~{Osterlind}, A.~{Dunkels}, J.~{Eriksson}, N.~{Finne}, and T.~{Voigt},
  ``Cross-level sensor network simulation with cooja,'' in \emph{Proceedings.
  2006 31st IEEE Conference on Local Computer Networks}, 2006, pp. 641--648.

\bibitem{ieee-tsch-std}
``{IEEE} standard for low-rate wireless networks,'' \emph{IEEE Std
  802.15.4-2015 (Revision of IEEE Std 802.15.4-2011)}, pp. 1--709, 2016.

\end{thebibliography}
\end{document}